\documentclass{PoS}

\newcommand{\be}{\begin{equation}}
\newcommand{\ee}{\end{equation}}

\newcommand{\reseteqnum}{\setcounter{equation}{0}}
\newcommand{\nn}{\nonumber}

\newcommand{\ovl}[1]{\overline{#1}}

\newcommand{\eqn}[1]{(\ref{#1})}

\newcommand{\bq}{{\overline{q}}}
\newcommand{\bu}{{\overline{u}}}
\newcommand{\bd}{{\overline{d}}}
\newcommand{\bs}{{\overline{s}}}

\newcommand{\vev}[1]{\left\langle #1 \right\rangle}

\newcommand{\pslash}{p\kern-1ex /}
\newcommand{\Dslash}{{\cal D}\kern-1.5ex /}

\title{Renormalization factor of four fermi operators with clover fermion and Iwasaki gauge action}

\ShortTitle{Renormalization factor of four fermi operators}

\author{\speaker{Yusuke Taniguchi}\\
Institute of Physics, University of Tsukuba,
Tsukuba, Ibaraki, 305-8571, Japan \\
        E-mail: \email{tanigchi@het.ph.tsukuba.ac.jp}}

\abstract{
Renormalization factors of four-quark operators are perturbatively
 calculated for the improved Wilson fermion with clover term and the
 Iwasaki gauge action.
 A main application shall be the $K\to\pi\pi$ decay amplitude and the
 calculation is restricted to the parity odd operator, for which
 the operators are multiplicatively renormalizable without mixing with
 wrong operators that have different chiral structures.
}

\FullConference{XXIX International Symposium on Lattice Field Theory \\
		 July 10 - 16 2011\\
		 Squaw Valley, Lake Tahoe, California}

\begin{document}

\section{Introduction}
Calculation of weak matrix elements of phenomenological interest
is one of major application of lattice QCD.
A calculation of four quark hadron matrix elements with the Wilson
fermion encounters an obstacle since unwanted mixing is introduced
through quantum correction with operators that have wrong chirality.

One of the solution is to make use of the parity odd operator.
By using discrete symmetries of the parity, the charge conjugation and
flavor exchanging transformations it was shown \cite{Donini:1999sf}
that the parity odd four quark operator has no extra mixing with wrong
operators even without chiral symmetry.
One of application of this virtue may be a calculation of the
$K\to\pi\pi$ decay amplitude with the Wilson fermion.

An improvement with the clover term is indivisible for the Wilson
fermion.
The RG improved gauge action of Iwasaki type has a good property
at lattice spacing around $a^{-1}\sim2$ GeV imitating that in the
continuum.
It is plausible to use a combination of the Iwasaki gauge action and the
improved Wilson fermion with clover term for our numerical simulation.
Unfortunately renormalization factors of the $\Delta S=1$ four quark
operators are not available for this combination of action except for
$\Delta S=2$ part \cite{Constantinou:2010zs}.
A purpose of this report is to give the renormalization factor of four
quark operators perturbatively which contribute to the $K\to\pi\pi$
decay.

\reseteqnum
\section{Four quark operators}
\label{sec:4fermi}

We adopt the Iwasaki gauge action and the improved Wilson fermion action
with the clover term.
The Feynman rules for this action is given in Ref.~\cite{Aoki:1998ar}.
We shall adopt the Feynman gauge and set the Wilson parameter $r=1$ in
the following.

We shall evaluate the renormalization factor of the following ten
operators
\begin{eqnarray}
&&
Q^{(2n-1)}=
\left(\bs d\right)_L\sum_{q=u,d,s}\alpha^{(n)}_q\left(\bq q\right)_L,
\;
Q^{(2n)}=\left(\bs\times d\right)_L
\sum_{q=u,d,s}\alpha^{(n)}_q\left(\bq\times q\right)_L,
\;
(n=1,2,5),
\\&&
Q^{(2n-1)}=
\left(\bs d\right)_L\sum_{q=u,d,s}\alpha^{(n)}_q\left(\bq q\right)_R,
\;
Q^{(2n)}=\left(\bs\times d\right)_L
\sum_{q=u,d,s}\alpha^{(n)}_q\left(\bq\times q\right)_R,
\;
(n=3,4),
\\&&
\alpha^{(1)}_q=\left(1,0,0\right),
\quad
\alpha^{(2)}_q=\alpha^{(3)}_q=\left(1,1,1\right),
\quad
\alpha^{(4)}_q=\alpha^{(5)}_q=\left(1,-\frac{1}{2},-\frac{1}{2}\right)
\end{eqnarray}
where
\begin{eqnarray}
\left(\bs d\right)_{R/L}=\bs\gamma_\mu\left({1\pm\gamma_5}\right)d
\end{eqnarray}
and $\times$ means a following contraction of the color indices
\begin{eqnarray}
Q^{(2)}=\left(\bs\times d\right)_L\left(\bu\times u\right)_L
=\left(\bs_a d_b\right)_L\left(\bu_b u_a\right)_L.
\end{eqnarray}

We are interested in the parity odd part only, which contribute to the
$K\to\pi\pi$ decay amplitude
\begin{eqnarray}
&&
 Q^{(2n-1)}_{VA+AV}=-Q^{(2n-1)}_{VA}-Q^{(2n-1)}_{AV},
\quad
 Q^{(2n)}_{VA+AV}=-Q^{(2n)}_{VA}-Q^{(2n)}_{AV},
\quad (n=1, 2, 5),
\\&&
 Q^{(2n-1)}_{VA-AV}=Q^{(2n-1)}_{VA}-Q^{(2n-1)}_{AV},
\quad
 Q^{(2n)}_{VA-AV}=Q^{(2n)}_{VA}-Q^{(2n)}_{AV},
\quad (n=3, 4),
\\&&
Q^{(2n-1)}_{VA}=
\left(\bs d\right)_V\sum_{q=u,d,s}\alpha^{(n)}_q\left(\bq q\right)_A,
\quad
Q^{(2n-1)}_{AV}=
\left(\bs d\right)_A\sum_{q=u,d,s}\alpha^{(n)}_q\left(\bq q\right)_V,
\\&&
Q^{(2n)}_{VA}=
\left(\bs\times d\right)_V
\sum_{q=u,d,s}\alpha^{(n)}_q\left(\bq\times q\right)_A,
\quad
Q^{(2n)}_{AV}=
\left(\bs\times d\right)_A
\sum_{q=u,d,s}\alpha^{(n)}_q\left(\bq\times q\right)_V,
\end{eqnarray}
where current-current vertex means
\begin{eqnarray}
\left(\bs d\right)_V\left(\bq q\right)_A
=\left(\bs\gamma_\mu d\right)\left(\bq\gamma_\mu\gamma_5q\right).
\end{eqnarray}

\reseteqnum
\section{Renormalization factor in $\ovl{\rm MS}$ scheme}

We renormalized the lattice bare operators $Q^{(k)}_{\rm lat}$ to get the
renormalized operator $Q^{(k)}_{\ovl{\rm MS}}$.
We adopt the $\ovl{\rm MS}$ scheme with DRED or NDR.
We notice there are two kinds of one loop corrections to the
operators.
One is given by gluon exchanging diagrams given in
Ref.~\cite{Constantinou:2010zs,Martinelli:1983ac} for $\Delta S=2$
operator and the other is the penguin diagrams given in
Ref.~\cite{Bernard:1987rw} for $\Delta S=1$ operators.

The renormalization of the operator is given by
\begin{eqnarray}
Q^{(i)}_{\ovl{\rm MS}}=Z_{ij}^gQ^{(j)}_{\rm lat}
+Z_i^{\rm pen}Q^{\rm pen}_{\rm lat}
+Z_i^{\rm sub}O^{\rm sub}_{\rm lat}
\end{eqnarray}
where $Q^{(j)}_{\rm lat}$ is the four quark operators on the lattice,
$Q^{\rm pen}_{\rm lat}$ is the QCD penguin operator and
$O^{\rm sub}_{\rm lat}$ is a lower dimensional operator to be
subtracted.
$Z_{ij}^g$ comes from gluon exchanging diagrams.
$Z_i^{\rm pen}$ is contribution from the penguin diagram.

\subsection{Gluon exchanging diagrams}

For gluon exchanging diagram the one loop contributions are evaluated
in terms of those to the quark bilinear operators by using the Fierz
rearrangement and the charge conjugation \cite{Martinelli:1983ac}.
Summing up contributions from three types of diagrams
\cite{Martinelli:1983ac} the one loop correction to the four quark
operators is given in a form 
\begin{eqnarray}
Q^{(i)}_{\rm one-loop}=T^{\rm lat}_{ij}Q^{(j)}_{\rm tree},
\end{eqnarray}
where $Q^{(j)}_{\rm tree}=Q^{(j)}_{VA\pm AV}$ is a tree level operator.
The correction factors are already evaluated for the improved action in
Ref.~\cite{Constantinou:2010zs} and is given as follows for our notation
of the four quark operators
\begin{eqnarray}
T^{\rm lat}_{11}&=&
T^{\rm lat}_{22}=T^{\rm lat}_{33}=T^{\rm lat}_{44}
=T^{\rm lat}_{99}=T^{\rm lat}_{10,10}
\nn\\&=&
\frac{g^2}{16\pi^2}
\left(-\frac{N^2+2}{N}\ln\left(\lambda a\right)^2
+\frac{N^2-2}{2N}\left(V_V+V_A\right)+\frac{1}{2N}\left(V_S+V_P\right)\right),
\\
T^{\rm lat}_{55}&=&
T^{\rm lat}_{77}
=\frac{g^2}{16\pi^2}
\left(-\frac{N^2-4}{N}\ln\left(\lambda a\right)^2
+\frac{N}{2}\left(V_V+V_A\right)
-\frac{1}{2N}\left(+V_S+V_P\right)
\right),
\\
T^{\rm lat}_{66}&=&
T^{\rm lat}_{88}
=\frac{g^2}{16\pi^2}
\left(-4\frac{N^2-1}{N}\ln\left(\lambda a\right)^2
+\frac{N^2-1}{2N}\left(V_S+V_P\right)\right),
\\
T^{\rm lat}_{12}&=&
T^{\rm lat}_{21}=T^{\rm lat}_{34}=T^{\rm lat}_{43}
=T^{\rm lat}_{9,10}=T^{\rm lat}_{10,9}
=\frac{g^2}{16\pi^2}\frac{1}{2}
\left(6\ln\left(\lambda a\right)^2
+V_V+V_A-V_S-V_P\right),
\\
T^{\rm lat}_{56}&=&
T^{\rm lat}_{78}
=\frac{g^2}{16\pi^2}\frac{1}{2}\left(
-6\ln\left(\lambda a\right)^2
-V_V-V_A+V_S+V_P\right),
\end{eqnarray}
where $\lambda$ is a gluon mass introduced for an infra red
regularization and the number of color is $N=3$.
$V_\Gamma$ is a finite part in one loop correction to the bilinear
operator, which is evaluated in Ref.~\cite{Aoki:1998ar} for various
gauge actions.

The renormalization factor is given by taking a ratio of quantum
corrections with that in the $\ovl{\rm MS}$ scheme multiplied with the
quark wave function renormalization factor $Z_2$
\begin{eqnarray}
&&
Z^g_{ii}(\mu a)=
\frac{\left(Z_2^{\ovl{\rm MS}}\right)^2\left(1+T_{ii}^{\ovl{\rm MS}}\right)}
{\left(Z_2^{\rm lat}\right)^2\left(1+T_{ii}^{\rm lat}\right)},
\\&&
Z^g_{ij}(\mu a)=T_{ij}^{\ovl{\rm MS}}-T_{ij}
\quad (i \neq j).
\end{eqnarray}
The correction factor in the DRED $\ovl{\rm MS}$ scheme is given by
\begin{eqnarray}
&&
 T^{\ovl{\rm MS}}_{11}=T^{\ovl{\rm MS}}_{22}
=T^{\ovl{\rm MS}}_{33}=T^{\ovl{\rm MS}}_{44}
=T^{\ovl{\rm MS}}_{99}=T^{\ovl{\rm MS}}_{10,10}
=\left(\frac{N^2+2}{N}\right)V^{\ovl{\rm MS}},
\\&&
 T^{\ovl{\rm MS}}_{12}=T^{\ovl{\rm MS}}_{21}
=T^{\ovl{\rm MS}}_{34}=T^{\ovl{\rm MS}}_{43}
=T^{\ovl{\rm MS}}_{9,10}=T^{\ovl{\rm MS}}_{10,9}
=-3V^{\ovl{\rm MS}},
\\&&
 T^{\ovl{\rm MS}}_{55}=T^{\ovl{\rm MS}}_{77}
=\left(\frac{N^2-4}{N}\right)V^{\ovl{\rm MS}},
\\&&
 T^{\ovl{\rm MS}}_{56}=T^{\ovl{\rm MS}}_{78}=3V^{\ovl{\rm MS}},
\\&&
 T^{\ovl{\rm MS}}_{66}=T^{\ovl{\rm MS}}_{88}
=4\frac{N^2-1}{N}V^{\ovl{\rm MS}},
\\&&
V^{\ovl{\rm MS}} = \frac{g^2}{16\pi^2}
\left(\log\left(\frac{\mu^2}{\lambda^2}\right)+1\right).
\end{eqnarray}
The same infra red regularization with the gluon mass should be adopted.
The quark wave function renormalization factor $Z_2$ is given in
Ref.~\cite{Aoki:1998ar}.

Substituting the above results we have
\begin{eqnarray}
Z^g_{11}(\mu a)&=&
Z^g_{22}(\mu a)=Z^g_{33}(\mu a)=Z^g_{44}(\mu a)=Z^g_{99}(\mu a)
=Z^g_{10,10}(\mu a)
\nn\\&=&
1+\frac{g^2}{16\pi^2}\left(\frac{3}{N}\ln\left(\mu a\right)^2+z^g_{11}\right),
\\
Z^g_{55}(\mu a)&=&Z^g_{77}(\mu a)
=1+\frac{g^2}{16\pi^2}\left(-\frac{3}{N}\ln\left(\mu a\right)^2+z^g_{55}\right),
\\
Z^g_{66}(\mu a)&=&Z^g_{88}(\mu a)
=1+\frac{g^2}{16\pi^2}\left(\frac{3\left(N^2-1\right)}{N}\ln\left(\mu a\right)^2
+z^g_{66}\right),
\\
Z^g_{12}(\mu a)&=&
Z^g_{21}(\mu a)=Z^g_{34}(\mu a)=Z^g_{43}(\mu a)=Z^g_{9,10}(\mu a)
=Z^g_{10,9}(\mu a)
\nn\\&=&
\frac{g^2}{16\pi^2}\left(-3\ln\left(\mu a\right)^2+z^g_{12}\right),
\\
Z^g_{56}(\mu a)&=&Z^g_{78}(\mu a)
=\frac{g^2}{16\pi^2}\left(3\ln\left(\mu a\right)^2+z^g_{56}\right),
\\
Z^g_{65}(\mu a)&=&Z^g_{87}(\mu a)=\frac{g^2}{16\pi^2}z^g_{65}=0.
\end{eqnarray}
The numerical value of the finite part is given in table
\ref{tab:gluon-exchanging} for $N=3$ as an expansion in $c_{\rm SW}$
\begin{eqnarray}
z^g_{ij}=z^{g(0)}_{ij}+c_{\rm SW}z^{g(1)}_{ij}+c_{\rm SW}^2z^{g(2)}_{ij}.
\end{eqnarray}
The finite part for the NDR scheme is given in table
\ref{tab:gluon-exchanging-NDR}.
We need to subtract the evanescent operators in the $\ovl{\rm MS}$
scheme, which comes from a difference of dimensionality from four for
gamma matrices in operator vertex.

\begin{table}[htb]
\caption{Finite part $z^g_{ij}$ of the renormalization factor from gluon
 exchanging diagrams in the DRED scheme.
}
\label{tab:gluon-exchanging}
\begin{center}
\begin{tabular}{|ccc|ccc|ccc|}
\hline
\multicolumn{3}{|c|}{$z^g_{11}$} &
\multicolumn{3}{c|}{$z^g_{55}$} &
\multicolumn{3}{c|}{$z^g_{66}$} \\
(0) & (1) & (2)  &
(0) & (1) & (2)  &
(0) & (1) & (2)  \\
\hline
-23.596 & 3.119 & 2.268 &
-25.183 & 5.420 & 2.923 &
-18.041 & -4.933 &-0.020 \\
\hline
\end{tabular}
\begin{tabular}{|ccc|ccc|}
\hline
\multicolumn{3}{|c|}{$z^g_{12}$} &
\multicolumn{3}{c|}{$z^g_{56}$} \\
(0) & (1) & (2)  &
(0) & (1) & (2)  \\
\hline
 -2.381 & 0.451 & -2.020 &
 2.381 & -0.451 & 2.020 \\
\hline
\end{tabular}
\end{center}
\end{table}
\begin{table}[htb]
\caption{Finite part $z^{g(0)}_{ij}$ of the renormalization factor from
 gluon exchanging diagrams in the NDR scheme.
 $c_{SW}$ dependent terms are the same as that in the DRED scheme
 $(n=1,2)$.}
\label{tab:gluon-exchanging-NDR}
\begin{center}
\begin{tabular}{|c|c|c|c|c|c|}
\hline
{$z^{g(0)}_{11}$} & {$z^{g(0)}_{55}$} & {$z^{g(0)}_{66}$} &
{$z^{g(0)}_{12}$} & {$z^{g(0)}_{56}$} & {$z^{g(0)}_{65}$} \\
\hline
$-24.096$ & $-25.350$ & $-19.708$ &
$ -4.881$ & $ -1.120$ & $-3$ \\
\hline
\end{tabular}
\end{center}
\end{table}

\subsection{Penguin diagrams}

Contribution from the penguin diagram is evaluated with the same
procedure as in Ref.~\cite{Bernard:1987rw} and the one loop correction
to the four quark operators is given in a form
\begin{eqnarray}
Q^{(i)}_{\rm one-loop}=\left(T^{\rm pen}_{i}\right)^{\rm lat}
Q^{\rm pen}_{\rm tree},
\end{eqnarray}
where $Q^{\rm pen}_{\rm tree}$ is the penguin operator at tree level
\begin{eqnarray}
Q^{\rm pen}=\left(Q^{(4)}_{VA+AV}+Q^{(6)}_{VA-AV}\right)
-\frac{1}{N}\left(Q^{(3)}_{VA+AV}+Q^{(5)}_{VA-AV}\right).
\end{eqnarray}

The correction factor is given by
\begin{eqnarray}
&&
\left(T^{\rm pen}_{i}\right)^{\rm lat}=\frac{g^2}{16\pi^2}\frac{C(Q_i)}{3}
\left(\ln{a^2p^2}+V_{\rm pen}^{\rm lat}\right)
\end{eqnarray}
with operator dependent factor
\begin{eqnarray}
&&
C\left(Q_1\right)=0,
\quad
C\left(Q_2\right)=1,
\quad
C\left(Q_3\right)=2,
\\&&
C\left(Q_4\right)=C\left(Q_6\right)=\sum_{q=u,d,s}=N_f,
\\&&
C\left(Q_5\right)=C\left(Q_7\right)=0,
\\&&
C\left(Q_8\right)=C\left(Q_{10}\right)=\sum_{q=u,d,s}\alpha_q=N_u-\frac{N_d}{2},
\\&&
C\left(Q_9\right)=-1.
\end{eqnarray}
$p$ is a momentum of intermediate gluon propagator given in terms of
external quark momentum, for which we set the on-shell condition.
The finite part is expanded as
\begin{eqnarray}
V_{\rm pen}^{\rm lat}=-1.7128+c_{\rm SW}\left(-1.0878\right).
\end{eqnarray}

The correction factor in the $\ovl{\rm MS}$ scheme is given in a similar
form
\begin{eqnarray}
&&
Q^{(i)}_{\rm one-loop}=\left(T^{\rm pen}_{i}\right)^{\ovl{\rm MS}}
Q^{\rm pen}_{\rm tree},
\\&&
\left(T^{\rm pen}_{i}\right)^{\ovl{\rm MS}}
=\frac{g^2}{16\pi^2}\frac{C(Q_i)}{3}
\left(\ln\left(\frac{p^2}{\mu^2}\right)-\frac{5}{3}-c\left(Q_i\right)\right)
\end{eqnarray}
With the same infra red regulator $p$.
The scheme dependent finite term is given by
\begin{eqnarray}
&&
c^{\rm (NDR)}\left(Q_2\right)=c^{\rm (NDR)}\left(Q_{2n-1}\right)=-1,
\quad
c^{\rm (NDR)}\left(Q_{2n}\right)=0,
\quad(n\ge2)
\\&&
c^{\rm (DRED)}\left(Q_2\right)=c^{\rm (DRED)}\left(Q_{2n-1}\right)
=c^{\rm (DRED)}\left(Q_{2n}\right)=\frac{1}{4},
\quad(n\ge2).
\end{eqnarray}

Combining these two contributions the renormalization factor for the
penguin operator is given by
\begin{eqnarray}
&&
Z_i^{\rm pen}=
\left(T^{\rm pen}_{i}\right)^{\ovl{\rm MS}}
-\left(T^{\rm pen}_{i}\right)^{\rm lat}
=\frac{g^2}{16\pi^{2}}\frac{C(Q_i)}{3}\left(-\ln a^2\mu^2
+z_i^{\rm pen}\right),
\\&&
z_i^{\rm pen}=-V_{\rm pen}^{\rm lat}-\frac{5}{3}-c_i.
\end{eqnarray}
Numerical value of the finite part is given in table \ref{tab:penguin}.
\begin{table}[htb]
\caption{Finite part of the renormalization factor from the penguin
 diagram.
 Coefficients of the term $c_{SW}^k (k=0,1)$ are given in the column 
 marked as $(k)$.}
\label{tab:penguin}
\begin{center}
\begin{tabular}{|l|lll|l|}
\hline
$z^{\rm pen}_i({\rm DRED})^{(0)}$ &
$z^{\rm pen}_2({\rm NDR})^{(0)}$ &
$z^{\rm pen}_{2n-1}({\rm NDR})^{(0)}$ &
$z^{\rm pen}_{2n}({\rm NDR})^{(0)}$ &
$(z^{\rm pen}_i)^{(1)}$ \\
\hline
$-0.2039$ &
 $1.0462$ &
 $1.0462$ &
 $0.0461$ &
 $1.0878$ \\
\hline
\end{tabular}
\end{center}
\end{table}

\subsection{Mixing with lower dimensional operator}

We shall evaluate the amputated quark bilinear vertex function given by
\begin{eqnarray}
I_{k;XY}^{\rm (sub)}&=&
\vev{Q^{(k)}_{XY} s_{a\alpha}(-p)\bd_{b\beta}(p)}_{\rm 1PI}.
\end{eqnarray}
We consider a leading contribution to the vertex at tree level, which
introduces mixing with lower dimensional operators.

We immediately get
\begin{eqnarray}
&&
I_{2n-1;VA}^{\rm (sub)}=-I_{2n-1;AV}^{\rm (sub)}=
\alpha^{(n)}_d\delta_{ab}\left(\gamma_5\right)_{\alpha\beta}
\left(I^{\rm (sub)}(m_d)-I^{\rm (sub)}(m_s)\right),
\\&&
I_{2n;VA}^{\rm (sub)}=-I_{2n;AV}^{\rm (sub)}=
N\alpha^{(n)}_d\delta_{ab}\left(\gamma_5\right)_{\alpha\beta}
\left(I^{\rm (sub)}(m_d)-I^{\rm (sub)}(m_s)\right),
\\&&
I^{\rm(sub)}(am)=\int\frac{d^4l}{(2\pi)^4}
\frac{4W(l,am)}{\sin^2l + W(l,am)^2},
\\&&
W(l,am)=am+\sum_{\mu}\left(1-\cos l_\mu\right).
\label{eqn:wilson-term}
\end{eqnarray}
which may be evaluated with an expansion in the quark mass
\begin{eqnarray}
I^{\rm (sub)}(m)&=&
\frac{1}{a^2}m\frac{d}{d(am)}I^{\rm (sub)}(0)
+\frac{1}{a}m^2\frac{1}{2}\frac{d^2}{d(am)^2}I^{\rm (sub)}(0)
+m^3\frac{1}{6}\frac{d^3}{d(am)^3}I^{\rm (sub)}(0)
+{\cal O}(a).
\nn\\
\end{eqnarray}
The numerical value is given by
\begin{eqnarray}
&&
\frac{d}{d(am)}I^{\rm (sub)}(0)=\frac{1}{16\pi^2}\left(-21.466\right),
\\&&
\frac{d^2}{d(am)^2}I^{\rm (sub)}(0)
=\frac{1}{16\pi^2}\left(-14.92\right).
\label{eqn:tree-level}
\end{eqnarray}

This contribution introduces a mixing with the lower dimensional
bilinear operator $(\bs\gamma_5d)$ multiplied with a mass difference
$(m_d-m_s)$.
As it is clear from \eqn{eqn:wilson-term} this is due to the chiral
symmetry breaking effect in the Wilson fermion.
It may be better not to expand in quark mass since the coefficient
\eqn{eqn:tree-level} is rather large and
${d^3}/{d(am)^3}I^{\rm (sub)}(0)$ term has an infra red divergence at
$m=0$.
The subtraction factor is given by
\begin{eqnarray}
&&
Z_{2n-1}^{\rm (sub)}=
-2\alpha^{(n)}_d\left(I^{\rm (sub)}(m_d)-I^{\rm (sub)}(m_s)\right),
\quad(n=3,4),
\\&&
Z_{2n}^{\rm (sub)}=
-2N\alpha^{(n)}_d\left(I^{\rm (sub)}(m_d)-I^{\rm (sub)}(m_s)\right),
\quad(n=3,4),
\\&&
Z_{2n-1}^{\rm (sub)}=Z_{2n}^{\rm (sub)}=0,\quad
(n=1,2,5).
\end{eqnarray}

\section{Conclusion}
\label{sec:concl}
In this report we have calculated the one-loop contributions
for the renormalization factors of parity odd four-quark operators,
which contribute to the $K\to\pi\pi$ decay amplitude,
in the improved Wilson fermion with clover term and the Iwasaki gauge
action.
The operators are multiplicatively renormalizable without any mixing
with wrong operators that have different chiral structures except for
the lower dimensional operator.

\section*{Acknowledgment}
This work is done for a collaboration with K.~-I.~Ishikawa, N.~Ishizuka,
A.~Ukawa and T.~Yoshi\'e.
This work is supported in part by Grants-in-Aid of the Ministry of
Education (Nos. 22540265, 23105701).

\end{document}